# The Dirac-Oscillator Green's function

A. D. Alhaidari

*Physics Department, King Fahd University of Petroleum & Minerals, Box 5047,
Dhahran 31261, Saudi Arabia*
e-mail: haidari@mailaps.org

We obtain the two-point Green's function for the relativistic Dirac-Oscillator problem. This is accomplished by setting up the relativistic problem in such a way that makes comparison with the nonrelativistic problem highly transparent and results in a map of the latter into the former. The relativistic bound states energy spectrum is obtained by locating the energy poles of this Green's function in a simple and straightforward manner.

## 1. INTRODUCTION

Recently, an effective approach has been developed for solving the Dirac equation for spherically symmetric local interactions. It was applied successfully to the solution of various relativistic problems [1-6]. These included, but not limited to, the Dirac-Coulomb, Dirac-Morse, Dirac-Scarf, Dirac-Pöschl-Teller, Dirac-Woods-Saxon, etc. The central idea in the approach is to separate the variables such that the two coupled first order differential equations resulting from the radial Dirac equation generate Schrödinger-like equations for the two spinor components. This makes the solution of the relativistic problem easily attainable by simple and direct correspondence with well-known exactly solvable nonrelativistic problems. The correspondence results in a parameter map that relates the relativistic to the nonrelativistic problem. Using this map and the known solutions (energy spectrum and wavefunctions) of the non-relativistic problem one can easily and directly obtain the relativistic spectrum and spinor wave-functions. The main objective in all previous applications of the approach was in obtaining the relativistic energy spectrum and the spinor wavefunctions [1-3,6]. In this article, however, we demonstrate how one can utilize the same approach in generating the two-point Green's function – an important object of prime significance to the calculation of relativistic physical processes. The main finding here is in obtaining the relativistic Green's function for the Dirac-Oscillator which, to the best of our knowledge, has not been calculated before. We start by setting up the relativistic problem then calculating the particle propagator in the Dirac-Oscillator potential.

Dirac equation is a relativistically covariant first order differential equation in 4-dimensional space-time for a four-component wavefunction ("spinor") $\psi$. For a free structureless particle it reads $\left(i\hbar\gamma^\mu\partial_\mu - mc\right)\psi = 0$, where $m$ is the rest mass of the particle and $c$ the speed of light. The summation convention over repeated indices is used. That is, $\gamma^\mu\partial_\mu \equiv \sum_{\mu=0}^{3}\gamma^\mu\partial_\mu = \gamma^0\partial_0 + \vec{\gamma}.\vec{\partial} = \gamma^0\frac{\partial}{c\partial t} + \vec{\gamma}.\vec{\nabla}$. $\{\gamma^\mu\}_{\mu=0}^{3}$ are four constant square matrices satisfying the anticommutation relation $\{\gamma^\mu,\gamma^\nu\} = \gamma^\mu\gamma^\nu + \gamma^\nu\gamma^\mu = 2\xi^{\mu\nu}$, where $\xi$ is the metric of Minkowski space-time which is equal to $\text{diag}(+,-,-,-)$. A four-dimensional matrix representation that satisfies this relation is as follows:

$$\gamma^0 = \begin{pmatrix} I & 0 \\ 0 & -I \end{pmatrix}, \quad \vec{\gamma} = \begin{pmatrix} 0 & \vec{\sigma} \\ -\vec{\sigma} & 0 \end{pmatrix} \tag{1}$$

where $I$ is the 2×2 unit matrix and $\vec{\sigma}$ are the three 2×2 hermitian Pauli matrices. In the atomic units $\hbar = m = 1$, the Compton wavelength $\lambdabar = \hbar/mc = 1/c$ and Dirac equation reads $\left(i\gamma^\mu \partial_\mu - \lambdabar^{-1}\right)\psi = 0$. Next, we let the Dirac spinor be coupled to the four component potential $A_\mu = (A_0, \vec{A})$. Gauge invariant coupling, which is accomplished by the "minimal" substitution $\partial_\mu \to \partial_\mu + iA_\mu$, transforms the free Dirac equation above to $\left[i\gamma^\mu(\partial_\mu + iA_\mu) - \lambdabar^{-1}\right]\psi = 0$ which, when written in details, reads as follows

$$i\lambdabar \tfrac{\partial}{\partial t}\psi = \left(-i\vec{\alpha}\cdot\vec{\nabla} + \vec{\alpha}\cdot\vec{A} + A_0 + \lambdabar^{-1}\beta\right)\psi \tag{2}$$

where $\vec{\alpha}$ and $\beta$ are the hermitian matrices

$$\vec{\alpha} = \gamma^0\vec{\gamma} = \begin{pmatrix} 0 & \vec{\sigma} \\ \vec{\sigma} & 0 \end{pmatrix} \text{ and } \beta = \gamma^0 = \begin{pmatrix} I & 0 \\ 0 & -I \end{pmatrix} \tag{3}$$

For time independent potentials, equation (2) gives the following matrix representation of the Dirac Hamiltonian (in units of $mc^2 = \lambdabar^{-2}$)

$$H = \begin{pmatrix} 1 + \lambdabar A_0 & -\lambdabar i\vec{\sigma}\cdot\vec{\nabla} + \lambdabar \vec{\sigma}\cdot\vec{A} \\ -\lambdabar i\vec{\sigma}\cdot\vec{\nabla} + \lambdabar \vec{\sigma}\cdot\vec{A} & -1 + \lambdabar A_0 \end{pmatrix} \tag{4}$$

Thus the eigenvalue wave equation reads $(H - \varepsilon)\psi = 0$, where $\varepsilon$ is the relativistic energy which is real and measured in units of $mc^2$.

Now, the space component $\vec{A}$ of the 4-potential could be eliminated by the usual gauge transformation $A_\mu \to A_\mu + \partial_\mu \eta$, $\psi \to e^{-i\lambdabar\eta}\psi$, where $\eta(\vec{r})$ is a real scalar function. Consequently, the contribution of the off-diagonal term $\lambdabar\vec{\sigma}\cdot\vec{A}$ could be eliminated ("gauged away") when the Dirac particle is coupled minimally to the potential $(A_0, \vec{A})$. However, in our approach the spinor is coupled in a non-minimal way to a four-vector potential $(A_0, \vec{A})$. This is accomplished by replacing the two off-diagonal terms $\vec{\sigma}\cdot\vec{A}$ in the above Hamiltonian (4) by $\pm i\vec{\sigma}\cdot\vec{A}$, respectively. That is, the Hamiltonian (4) is replaced by the following

$$H = \begin{pmatrix} 1 + \lambdabar A_0 & -i\lambdabar\vec{\sigma}\cdot\vec{\nabla} + i\lambdabar\vec{\sigma}\cdot\vec{A} \\ -i\lambdabar\vec{\sigma}\cdot\vec{\nabla} - i\lambdabar\vec{\sigma}\cdot\vec{A} & -1 + \lambdabar A_0 \end{pmatrix} \tag{5}$$

It should be noted that this type of coupling does not support an interpretation of $(A_0, \vec{A})$ as the electromagnetic potential unless, of course, $\vec{A} = 0$ (e.g., the Coulomb potential). Likewise, $H$ does not have local gauge symmetry. That is, the associated wave equation is not invariant under the electromagnetic gauge transformation mentioned above. The hermitian (real) Lagrangian that gives the gauge invariant Dirac equation (2) with minimal coupling to the electromagnetic potential $A_\mu$ is

$$\int \left(\tfrac{i}{2}\bar{\psi}\overleftrightarrow{\partial}\psi - \lambdabar^{-1}\bar{\psi}\psi - \bar{\psi}\slashed{A}\psi\right) d^4x \tag{6}$$

where $\not{C} \equiv \gamma^\mu C_\mu = \begin{pmatrix} C_0 & \vec{\sigma}\cdot\vec{C} \\ -\vec{\sigma}\cdot\vec{C} & -C_0 \end{pmatrix}$ and $\vec{\not{\partial}} = \gamma^\mu(\vec{\partial}_\mu - \overleftarrow{\partial}_\mu)$. On the other hand, the Dirac equation with nonminimal coupling to the potential $A_\mu$ is obtained from the real Lagrangian

$$\int \left( \tfrac{i}{2} \bar{\psi}\vec{\not{\partial}}\psi - \lambdabar^{-1}\bar{\psi}\psi - \bar{\psi}\slashed{A}\psi \right) d^4x \qquad (7)$$

where $\slashed{A} \equiv \gamma^0 C_0 + i\vec{\alpha}\cdot\vec{C} = \begin{pmatrix} C_0 & i\vec{\sigma}\cdot\vec{C} \\ i\vec{\sigma}\cdot\vec{C} & -C_0 \end{pmatrix}$. It is to be noted that relativistic invariance of the term $\bar{\psi}\slashed{A}\psi$ in (6) dictates that the electromagnetic potential $A_\mu$ transforms like a 4-vector. That is $A'_\mu = \Lambda_\mu{}^\nu A_\nu$, where $\Lambda$ is the six-parameter 4×4 Lorentz transformation matrix satisfying $\Lambda^\top \xi \Lambda = \xi$. On the other hand, relativistic invariance of the term $\bar{\psi}\slashed{A}\psi$ in (7) gives the transformation property of the 4-potential $A_\mu$ which does not transform like a 4-vector.

Now, we choose a reference frame where $A_0 = 0$ and impose spherical symmetry by taking $\vec{A} = \hat{r}W(r)$, where $\hat{r}$ is the radial unit vector and $W(r)$ is a real radial potential function. In this case, the angular variables could be separated and we can write the spinor wavefunction as [7]

$$\psi = \begin{pmatrix} i[\phi^+(r)/r]\chi^j_{\ell m} \\ [\phi^-(r)/r]\vec{\sigma}\cdot\hat{r}\chi^j_{\ell m} \end{pmatrix} \qquad (8)$$

where $\phi^\pm$ are real radial functions and the angular wavefunction for the two-component spinor is written as

$$\chi^j_{\ell m}(\hat{r}) = \frac{1}{\sqrt{2\ell+1}} \begin{pmatrix} \sqrt{\ell \pm m + 1/2}\; Y_\ell^{m-1/2} \\ \pm\sqrt{\ell \mp m + 1/2}\; Y_\ell^{m+1/2} \end{pmatrix}, \qquad \text{for } j = \ell \pm \tfrac{1}{2} \qquad (9)$$

$Y_\ell^{m\pm 1/2}$ is the spherical harmonic function and $m = -j, -j+1, \ldots, j$. Spherical symmetry gives $i\vec{\sigma}\cdot(\vec{r}\times\vec{\nabla})\psi(r,\hat{r}) = -(1+\kappa)\psi(r,\hat{r})$, where $\kappa$ is the spin-orbit quantum number defined as $\kappa = \pm(j+\tfrac{1}{2}) = \pm 1, \pm 2, \ldots$ for $\ell = j \pm \tfrac{1}{2}$. Using this we obtain the following useful relations

$$\begin{aligned}
(\vec{\sigma}\cdot\vec{\nabla})(\vec{\sigma}\cdot\hat{r})F(r)\chi^j_{\ell m} &= \left( \frac{dF}{dr} + \frac{1-\kappa}{r}F \right)\chi^j_{\ell m} \\
(\vec{\sigma}\cdot\vec{\nabla})F(r)\chi^j_{\ell m} &= \left( \frac{dF}{dr} + \frac{1+\kappa}{r}F \right)(\vec{\sigma}\cdot\hat{r})\chi^j_{\ell m}
\end{aligned} \qquad (10)$$

Employing these in the wave equation $(H-\varepsilon)\psi = 0$ results in the following 2×2 matrix equation for the two radial spinor components

$$\begin{pmatrix} 1-\varepsilon & \lambdabar\left[\frac{\kappa}{r} + W(r) - \frac{d}{dr}\right] \\ \lambdabar\left[\frac{\kappa}{r} + W(r) + \frac{d}{dr}\right] & -1-\varepsilon \end{pmatrix} \begin{pmatrix} \phi^+(r) \\ \phi^-(r) \end{pmatrix} = 0 \qquad (11)$$

This gives the following equation for one spinor component in terms of the other

$$\phi^\mp(r) = \frac{\lambdabar}{\varepsilon \pm 1}\left[ \frac{\kappa}{r} + W(r) \pm \frac{d}{dr} \right]\phi^\pm(r) \qquad (12)$$

On the other hand, the resulting Schrödinger-like wave equation for the two spinor components reads

$$\left[-\frac{d^2}{dr^2}+\frac{\kappa(\kappa\pm 1)}{r^2}+W^2+2\kappa\frac{W}{r}\mp\frac{dW}{dr}-\frac{\varepsilon^2-1}{\hbar^2}\right]\phi^\pm=0 \qquad (13)$$

The objective of adding a potential, which is linear in the coordinate, to the Dirac equation in an analogy to the kinetic energy term which is linear in the momentum lead Moshinsky and Szczepaniak to the solution of the Dirac-Oscillator problem [8]. The nonrelativistic limit reproduces the usual Harmonic oscillator. The linear potential had to be added to the odd part of the Dirac operator resulting in a potential coupling which is a special case of that given in the Hamiltonian of Eq. (5) above. Subsequently, the Dirac-Oscillator attracted a lot of attention in the literature [9]. Our contribution here is to find its two-point Green's function using the tools of the approach mentioned above. In this setting, the Dirac-Oscillator is the system described by Eq. (11) with $W(r) = \omega^2 r$, where $\omega$ is the oscillator frequency.

## 2. DIRAC-OSCILLATOR GREEN'S FUNCTION

The relativistic 4×4 two-point Green's function $G(\vec{r},\vec{r}',\varepsilon)$ satisfies the inhomogeneous matrix wave equation $(H-\varepsilon)G=-\hbar^2\delta(\vec{r}-\vec{r}')$, where the energy $\varepsilon$ does not belong to the spectrum of $H$. For problems with spherical symmetry, the 2×2 radial component $\mathcal{G}_\kappa(r,r',\varepsilon)$ of $G$ satisfies $(H_\kappa-\varepsilon)\mathcal{G}_\kappa=-\hbar^2\delta(r-r')$, where $H_\kappa$ is the radial Hamiltonian operator in Eq. (11). It should be noted that our definition of the radial component of the Green's function differs by a factor of $(rr')^{-1}$ from other typical definitions. We write $\mathcal{G}_\kappa$ as

$$\mathcal{G}_\kappa(r,r',\varepsilon)=\begin{pmatrix}\mathcal{G}_\kappa^{++} & \mathcal{G}_\kappa^{+-} \\ \mathcal{G}_\kappa^{-+} & \mathcal{G}_\kappa^{--}\end{pmatrix} \qquad (14)$$

where $\mathcal{G}_\kappa(r,r',\varepsilon)^\dagger=\mathcal{G}_\kappa(r',r,\varepsilon)$. The equations satisfied by the elements of $\mathcal{G}_\kappa$ are obtained from $(H_\kappa-\varepsilon)\mathcal{G}_\kappa=-\hbar^2\delta(r-r')$. They parallel Eq. (12) and Eq. (13) for $\phi^\pm$ and read as follows:

$$\left[-\frac{d^2}{dr^2}+\frac{\kappa(\kappa\pm 1)}{r^2}+\omega^4 r^2+\omega^2(2\kappa\mp 1)-\frac{\varepsilon^2-1}{\hbar^2}\right]\mathcal{G}_\kappa^{\pm\pm}(r,r',\varepsilon)=-(1\pm\varepsilon)\delta(r-r') \qquad (15)$$

$$\mathcal{G}_\kappa^{\mp\pm}(r,r',\varepsilon)=\frac{\hbar}{\varepsilon\pm 1}\left[\frac{\kappa}{r}+\omega^2 r\pm\frac{d}{dr}\right]\mathcal{G}_\kappa^{\pm\pm}(r,r',\varepsilon) \qquad (16)$$

We compare Eq. (15) to that of the nonrelativistic radial Green's function $g_\ell(r,r',E)$ for the three dimensional isotropic oscillator:

$$\left[-\frac{d^2}{dr^2}+\frac{\ell(\ell+1)}{r^2}+\omega^4 r^2-2E\right]g_\ell(r,r',E)=-2\delta(r-r') \qquad (17)$$

where $\ell$ is the angular momentum quantum number and $E$ is the nonrelativistic energy. The comparison gives the following two maps between the relativistic and non-relativistic problems. The map concerning $\mathcal{G}_\kappa^{++}$ is

$$g_\ell\to 2\mathcal{G}_\kappa^{++}/(1+\varepsilon),\quad \ell\to\begin{cases}\kappa, & \kappa>0 \\ -\kappa-1, & \kappa<0\end{cases},\quad E\to(\varepsilon^2-1)/2\hbar^2-\omega^2(\kappa-1/2) \qquad (18)$$

The choice $\ell \to \kappa$ or $\ell \to -\kappa-1$ depends on whether $\kappa > 0$ or $\kappa < 0$, respectively. On the other hand, the map for $\mathcal{G}_\kappa^{--}$ is as follows:

$$g_\ell \to 2\mathcal{G}_\kappa^{--}/(1-\varepsilon), \quad \ell \to \begin{cases} \kappa-1, \kappa>0 \\ -\kappa, \kappa<0 \end{cases}, \quad E \to (\varepsilon^2-1)/2\lambdabar^2 - \omega^2(\kappa+1/2) \quad (19)$$

Similarly, the choice $\ell \to \kappa-1$ or $\ell \to -\kappa$ depends on whether $\kappa$ is positive or negative, respectively. Now, the nonrelativistic radial Green's function for the harmonic oscillator is well known [10]. It could be written as

$$g_\ell(r,r',E) = \frac{\Gamma(\frac{2\ell+3}{4} - E/2\omega^2)}{\omega^2 \Gamma(\ell+3/2)} \frac{1}{\sqrt{rr'}} \mathcal{M}_{E/2\omega^2, \frac{2\ell+1}{4}}(\omega^2 r_<^2) \mathcal{W}_{E/2\omega^2, \frac{2\ell+1}{4}}(\omega^2 r_>^2) \quad (20)$$

where $\Gamma$ is the gamma function, $\mathcal{M}_{a,b}$ and $\mathcal{W}_{a,b}$ are the Whittaker functions of the first and second kind, respectively [11]. $r_<(r_>)$ is the smaller (larger) of $r$ and $r'$. The two mappings (18) and (19) transform this nonrelativistic Green's function into the following solutions of Eq. (15):

$$\mathcal{G}_\kappa^{++} = \frac{1+\varepsilon}{2\omega^2} \frac{1}{\sqrt{rr'}} \begin{cases} \frac{\Gamma(-\mu+2\nu)}{\Gamma(2\nu+1)} \mathcal{M}_{\mu-\nu+\frac{1}{2},\nu}(\omega^2 r_<^2) \mathcal{W}_{\mu-\nu+\frac{1}{2},\nu}(\omega^2 r_>^2), & \kappa>0 \\ \frac{\Gamma(-\mu)}{\Gamma(-2\nu+1)} \mathcal{M}_{\mu-\nu+\frac{1}{2},-\nu}(\omega^2 r_<^2) \mathcal{W}_{\mu-\nu+\frac{1}{2},-\nu}(\omega^2 r_>^2), & \kappa<0 \end{cases} \quad (21)$$

$$\mathcal{G}_\kappa^{--} = \frac{1-\varepsilon}{2\omega^2} \frac{1}{\sqrt{rr'}} \begin{cases} \frac{\Gamma(-\mu+2\nu)}{\Gamma(2\nu)} \mathcal{M}_{\mu-\nu,\nu-\frac{1}{2}}(\omega^2 r_<^2) \mathcal{W}_{\mu-\nu,\nu-\frac{1}{2}}(\omega^2 r_>^2), & \kappa>0 \\ \frac{\Gamma(-\mu+1)}{\Gamma(-2\nu+2)} \mathcal{M}_{\mu-\nu,-\nu+\frac{1}{2}}(\omega^2 r_<^2) \mathcal{W}_{\mu-\nu,-\nu+\frac{1}{2}}(\omega^2 r_>^2), & \kappa<0 \end{cases} \quad (22)$$

where $\mu = (\varepsilon^2-1)/4\lambdabar^2\omega^2$ and $\nu = (\kappa+\frac{1}{2})/2$. The off-diagonal elements of $\mathcal{G}_\kappa$ are obtained by substituting these in Eq. (16), which could be rewritten as

$$\mathcal{G}_\kappa^{\mp\pm}(r,r',\varepsilon) = \frac{\lambdabar}{\varepsilon\pm 1} \frac{1}{\sqrt{rr'}} \left[ \frac{\kappa\mp\frac{1}{2}}{r} + \omega^2 r \pm \frac{d}{dr} \right] \sqrt{rr'}\, \mathcal{G}_\kappa^{\pm\pm}(r,r',\varepsilon) \quad (23)$$

Using the differential properties of the Whittaker functions [11] we obtain relations (A1) and (A2) in the Appendix, which when used in Eq. (23) give

$$\mathcal{G}_\kappa^{-+}(r,r',\varepsilon) = \mathcal{G}_\kappa^{+-}(r',r,\varepsilon) = \frac{\Gamma(-\mu+2\nu)}{\Gamma(2\nu)} \frac{\lambdabar/\omega}{\sqrt{rr'}}$$
$$\times \Big[ \theta(r'-r)\mathcal{M}_{\mu-\nu,\nu-\frac{1}{2}}(\omega^2 r^2) \mathcal{W}_{\mu-\nu+\frac{1}{2},\nu}(\omega^2 r'^2)$$
$$+ \frac{\mu}{2\nu}\theta(r-r')\mathcal{M}_{\mu-\nu+\frac{1}{2},\nu}(\omega^2 r'^2) \mathcal{W}_{\mu-\nu,\nu-\frac{1}{2}}(\omega^2 r^2) \Big] \quad , \kappa>0 \quad (24a)$$

$$\mathcal{G}_\kappa^{-+}(r,r',\varepsilon) = \mathcal{G}_\kappa^{+-}(r',r,\varepsilon) = \frac{\Gamma(-\mu+1)}{\Gamma(-2\nu+2)} \frac{\lambdabar/\omega}{\sqrt{rr'}}$$
$$\times \Big[ \theta(r'-r)\mathcal{M}_{\mu-\nu,-\nu+\frac{1}{2}}(\omega^2 r^2) \mathcal{W}_{\mu-\nu+\frac{1}{2},-\nu}(\omega^2 r'^2)$$
$$+ (2\nu-1)\theta(r-r')\mathcal{M}_{\mu-\nu+\frac{1}{2},-\nu}(\omega^2 r'^2) \mathcal{W}_{\mu-\nu,-\nu+\frac{1}{2}}(\omega^2 r^2) \Big], \kappa<0 \quad (24b)$$

It is worth noting (although might be obvious) that in the nonrelativistic limit ($\lambdabar \to 0$, $\varepsilon \to 1+\lambdabar^2 E$) the off-diagonal elements go to the limit like $\lambdabar$, whereas the lower diagonal element $\mathcal{G}_\kappa^{--}$ goes like $\lambdabar^2$.

Finally, one can easily verify that the relativistic bound states energy spectrum of the Dirac-Oscillator [5,9] is located at the energy poles of these components of the Green's function. It is simply obtained by taking the argument of the gamma function in the numerator to be equal to $-n$, where $n = 0,1,2\ldots$ That is by taking $-\mu+2\nu = -n$ for $\kappa > 0$ and $-\mu = -n$ for $\kappa < 0$ giving:

$$\varepsilon_n = \begin{cases} \pm\sqrt{1+4\hbar^2\omega^2(n+\kappa+\frac{1}{2})} &, \kappa > 0 \\ \pm\sqrt{1+4\hbar^2\omega^2 n} &, \kappa < 0 \end{cases} \qquad (26)$$

## ACKNOWLEDGMENTS

The author is grateful to Dr. M. I. Al-Suwaiyel and KACST Library for the valuable support in literature survey.

## APPENDIX

The following are some useful relations which could be obtained by using the differential formulas and recurrence relations of the Whittaker functions [11]:

$$\left(\frac{d}{dx} + \frac{2b-1}{x} \pm x\right)\mathcal{M}_{a,b}(x^2) = 4b\,\mathcal{M}_{a\mp\frac{1}{2},b-\frac{1}{2}}(x^2)$$
$$\left(\frac{d}{dx} - \frac{2b+1}{x} \pm x\right)\mathcal{M}_{a,b}(x^2) = \left(\pm 1 - \frac{a}{b+\frac{1}{2}}\right)\mathcal{M}_{a\mp\frac{1}{2},b+\frac{1}{2}}(x^2) \qquad (A1)$$

$$\left(\frac{d}{dx} - \frac{1\pm 2b}{x} + x\right)\mathcal{W}_{a,b}(x^2) = 2(a \mp b - \tfrac{1}{2})\mathcal{W}_{a-\frac{1}{2},b\pm\frac{1}{2}}(x^2)$$
$$\left(\frac{d}{dx} - \frac{1\pm 2b}{x} - x\right)\mathcal{W}_{a,b}(x^2) = -2\mathcal{W}_{a+\frac{1}{2},b\pm\frac{1}{2}}(x^2) \qquad (A2)$$

$$\left(\frac{d}{dx} + \frac{b-\frac{1}{2}}{x} - \frac{a}{2b-1}\right)\mathcal{M}_{a,b}(x) = 2b\,\mathcal{M}_{a,b-1}(x)$$
$$\left(\frac{d}{dx} - \frac{b+\frac{1}{2}}{x} + \frac{a}{2b+1}\right)\mathcal{M}_{a,b}(x) = \frac{1/8}{b+1}\left[1-\left(\frac{a}{b+\frac{1}{2}}\right)^2\right]\mathcal{M}_{a,b+1}(x) \qquad (A3)$$

$$\left(\frac{d}{dx} + \frac{b-\frac{1}{2}}{x} - \frac{a}{2b-1}\right)\mathcal{W}_{a,b}(x) = -\frac{1}{2}\left(1+\frac{a}{b-\frac{1}{2}}\right)\mathcal{W}_{a,b-1}(x)$$
$$\left(\frac{d}{dx} - \frac{b+\frac{1}{2}}{x} + \frac{a}{2b+1}\right)\mathcal{W}_{a,b}(x) = \frac{1}{2}\left(-1+\frac{a}{b+\frac{1}{2}}\right)\mathcal{W}_{a,b+1}(x) \qquad (A4)$$

# REFERENCES

[1] A. D. Alhaidari, Phys. Rev. Lett. **87**, 210405 (2001); **88**, 189901 (2002)

[2] A. D. Alhaidari, J. Phys. A **34**, 9827 (2001); **35**, 6207 (2002)

[3] A. D. Alhaidari, Int. J. Mod. Phys. A **17**, 4551 (2002)

[4] A. D. Alhaidari, Phys. Rev. A **65**, 042109 (2002); **66**, 019902 (2002)

[5] A. D. Alhaidari, Int. J. Mod. Phys. A **18**, 4955 (2003)

[6] J-Y Guo, X. Z. Fang, and F-X Xu, Phys. Rev. A **66**, 062105 (2002); J-Y Guo, J. Meng, and F-X Xu, Chin. Phys. Lett. **20**, 602 (2003)

[7] J. D. Bjorken and S. D. Drell, *Relativistic Quantum Mechanics* (McGraw Hill, New York, 1965)

[8] M. Moshinsky and A. Szczepaniak, J. Phys. A **22**, L817 (1989)

[9] See, for example, J. Bentez, R. P. Martinez-y-Romero, H. N. Nunez-Yepez, and A. L. Salas-Brito, Phys. Rev. Lett. **64**, 1643 (1990); O. L. de Lange, J. Phys. A **24**, 667 (1991); V. M. Villalba, Phys. Rev. A **49**, 586 (1994); P. Rozmej and R. Arvieu, J. Phys. A **32**, 5367 (1999); R. Szmytkowski and M. Gruchowski, J. Phys. A **34**, 4991 (2001)

[10] See, for examples, J. Bellandi Filho and E. S. Caetano Neto, Lett. Al Nuovo Cimento **16**, 331 (1976); E. Capelas De Oliveira, Revista Bras. de Fisica **9**, 697 (1979); J. H. Macek, S. Yu. Ovchinnikov and D. B. Khrebtukov, Rad. Phys. Chem. **59**, 149 (2000)

[11] W. Magnus, F. Oberhettinger, and R. P. Soni, *Formulas and Theorems for the Special Functions of Mathematical Physics*, 3$^{rd}$ edition (Springer-Verlag, New York, 1966); H. Buchholz, *The Confluent Hypergeometric Function* (Springer-Verlag, New York, 1969); I. S. Gradshtein and I. M. Ryzhik, *Table of Integrals, Series and Products* (Academic, New York, 1980); H. Bateman and A. Erdélyi, *Higher Transcendental Functions* (McGraw-Hill, New York, 1953).